\documentclass[aps,reprint,nofootinbib,nobibnotes,notitlepage,superscriptaddress,twocolumn,prl,
 amsmath,amssymb
]{revtex4-2}

\usepackage[caption=false]{subfig}
\usepackage{braket}
\usepackage{float}
\usepackage{lipsum}
\usepackage{graphicx}
\usepackage{dcolumn}
\usepackage{bm}
\usepackage{natbib}
\usepackage{hyperref}
\hypersetup{
	colorlinks = true,
    linkcolor = Red,
    urlcolor  = Red,
    citecolor = Red
}
\usepackage{subfig}
\usepackage{microtype}

\usepackage{verbatim}
\usepackage[amssymb]{SIunits}
\usepackage{tabularx}
\usepackage[dvipsnames]{xcolor}
\usepackage{wasysym}
\usepackage[export]{adjustbox}
\usepackage{multirow}
\usepackage{tikz}

\def\be{\begin{equation}}
\def\ee{\end{equation}}
\usepackage{enumitem}

\usepackage{color}
\definecolor{darkgreen}{RGB}{0,120,0}

\definecolor{darkgreen}{RGB}{0,120,0}
\newcommand{\resub}[1]{{#1}}

\newcommand{\delD}[1]{(2\pi)^3\delta_\mathrm{D}\left({#1}\right)}

\newcommand{\av}[1]{\left\langle{#1}\right\rangle} 

\newcommand{\vk}{\vec k}
\newcommand{\hk}{\hat{\vec k}}

\newcommand{\Si}{\mathsf{S}^{-1}}
\newcommand{\F}{\mathcal{F}}

\newcommand{\hr}{\hat{\vec r}}

\renewcommand{\vr}{\vec r}

\renewcommand{\P}{\mathcal{P}}

\def\beq{\begin{eqnarray}}
\def\eeq{\end{eqnarray}}
\let\vec\mathbf

\usepackage{empheq}

\usepackage{booktabs}

\makeatletter 
    
\renewcommand\onecolumngrid{
\do@columngrid{one}{\@ne}%
\def\set@footnotewidth{\onecolumngrid}
\def\footnoterule{\kern-6pt\hrule width 1.5in\kern6pt}%
}

\renewcommand\twocolumngrid{
        \def\footnoterule{
        \dimen@\skip\footins\divide\dimen@\thr@@
        \kern-\dimen@\hrule width.5in\kern\dimen@}
        \do@columngrid{mlt}{\tw@}
}%

\makeatother    
\begin{document}

\title{What Shape is the Inflationary Bispectrum?}

\author{Oliver~H.\,E.~Philcox}
\email{ohep2@cantab.ac.uk}
\affiliation{Leinweber Institute for Theoretical Physics at Stanford, 382 Via Pueblo, Stanford, CA 94305, USA}
\affiliation{Kavli Institute for Particle Astrophysics and Cosmology, 382 Via Pueblo, Stanford, CA 94305, USA}

\begin{abstract} 
\noindent Non-linear interactions during inflation generate non-Gaussianities in the distribution of primordial curvature. In many theories, the physics is \textit{scale-invariant}, such that the induced three-point function depends solely on a dimensionless shape function $S(x,y)\sim k^6B_\zeta(kx,ky,k)$. To confront such models with observations, one typically builds specialized estimators for each shape, then applies them to cosmic microwave background datasets at significant computational expense. In this \textit{Letter}, we take a different approach, directly reconstructing $S(x,y)$ from observations using an efficient logarithmically-binned estimator in primordial-space (motivated by the modal program). Applying this to temperature and polarization maps from \textit{Planck}, we obtain high-resolution shape measurements across the full $(x,y)$-plane, including squeezed limits. Our approach is close-to-optimal, highly interpretable, and preserves the information content on (optimally-analyzed) standard templates within $\approx 10\%$; moreover, we can use it to assess the scale-dependence of our constraints, finding that \textit{Planck} is sensitive to $\approx 6$ $e$-folds of non-Gaussian evolution with a peak sensitivity around $0.1h\,\mathrm{Mpc}^{-1}$. Since we work directly in shape-space, data and theory can be compared in milliseconds. As an example, we perform a search for massive particle exchange using a suite of over $20\,000$ theoretical templates computed with exact bootstrap methods across a wide range of masses, spins, and sound-speeds; the spin-two analysis yields a maximum significance of $2.6\sigma$. Our approach can be used to probe a wide range of scale-invariant models in orders-of-magnitude less time than with direct estimators, \resub{facilitating new exploration of the inflationary paradigm.}
\end{abstract}

\maketitle

\noindent Observations of the Cosmic Microwave Background are consistent with an inflationary primordial epoch, featuring an almost-exponential background expansion combined with quantum fluctuations around a Bunch Davies vacuum \citep[e.g.,][]{WMAP:2003elm,Planck:2018jri}. A central prediction of this model is a nearly \textit{scale-invariant} spectrum of primordial curvature fluctuations, defined by the two-point function $P_\zeta(k)\propto k^{n_s-4}$ where $n_s$ is close to unity \citep{Guth:1980zm,Linde:1981mu,Mukhanov:1990me,Mukhanov:1981xt,Bunch:1978yq}. Practically, this implies that the background geometry of the Universe is almost de Sitter, with corrections suppressed by slow-roll parameters. Asserting that the inflationary Lagrangian also (almost) obeys de Sitter isometries greatly restricts the space of inflationary models, for example ensuring that masses and coupling coefficients are time-invariant. Importantly, this assumption implies that all inflationary correlators are approximately scale-invariant, with, for example, the bispectrum, $B_\zeta$ scaling as $\alpha^{-6}$ under the dilation $k\to \alpha k$. For this reason, it is common to define a dimensionless shape function $S(x,y)\sim k^6(xy)^2B_\zeta(kx,ky,k)$ which is a function only of momentum ratios \citep{Maldacena:2002vr,Babich:2004gb,2009PhRvD..80d3510F}. 

A wide variety of scale-invariant inflationary models have been proposed, involving a plethora of interactions, particles, currents, and beyond. Many of these models are unified in the Effective Field Theory of Inflation \citep{Weinberg:2008hq,Cheung:2007st,Senatore:2010wk,Senatore:2009gt} (see also \citep{Chen:2006nt,Chen:2008wn}), which categorizes all possible low-order interactions of a set of inflationary fields consistent with the assumed symmetries of the system. Due to scale invariance, these models cannot be distinguished by measurements of the two-point function alone (at least in the absence of gravitational wave detections), thus our chief diagnostic tool is the three-point function, $S(x,y)$, whose structure encodes the field content and interactions, and whose amplitude describes the inflationary couplings. 

A particularly interesting example is the primordial exchange of massive particles, which leads to oscillatory signals in the squeezed limit of $S(x,y)$ (with $x\ll y\approx 1$), whose frequency and angular dependence encode the mass and spin of the exchange particle respectively \citep[e.g.,][]{Chen:2009we,Chen:2009zp,Lee:2016vti,Arkani-Hamed:2015bza,Chen:2015lza}; this facilitates particle spectrometry for masses around the inflationary Hubble scale (which could exceed $10^{13}\,\mathrm{GeV}$ \citep{BICEPKeck:2022mhb}). Despite the simple form of the correlator in the squeezed limit, it has become clear that a full analysis of the exchange interactions requires going beyond the squeezed limits, and analyzing the full momentum dependence of the correlator \citep[e.g.,][]{Pimentel:2022fsc,Jazayeri:2022kjy,Werth:2023pfl,Lee:2016vti}.

If new physics is lurking in the inflationary three-point function, how can we detect it? To date, most analyses of primordial non-Gaussianity (PNG) focus on the CMB temperature and polarization anisotropies, which currently dominate the inflationary information content \citep[e.g.,][]{Senatore:2009gt,Planck:2019kim,Jung:2025nss,Sohn:2024xzd,Meerburg:2019qqi,Creminelli:2005hu}, though large-scale structure is beginning to yield competitive results \citep{Chudaykin:2025vdh}. Despite its linear nature, the CMB is difficult to analyze, owing to its high dimensionality ($\mathcal{O}(10^7)$ pixels) and oscillatory transfer functions, which relate the observed anisotropies to primordial curvature \citep[e.g.,][]{Komatsu:2003iq,Komatsu:2001rj}. Typically, one searches for inflationary signatures by performing targeted searches for specific models of interest, rather than attempting to measure a set of model-independent summary statistics that can then be compared to theoretical models \citep[e.g.,][]{Planck:2019kim}.
Many of the strongest constraints on inflationary models arise from Komatsu-Spergel-Wandelt estimators \citep{Komatsu:2001wu,Komatsu:2003iq,Komatsu:2001rj,2011MNRAS.417....2S}, which compress the full set of $\mathcal{O}(10^{21})$ bispectrum configurations into a single number. Although quasi-optimal, this procedure is computationally demanding and must be repeated for every primordial bispectrum of interest. Furthermore, it has limited interpretability, and it remains difficult to assess if there is some residual pattern lurking in the data.

Several methods have been proposed to analyze the CMB three-point function in a more generic fashion, including via computation of the binned harmonic-space bispectrum \citep{Bucher:2015ura,Philcox:2023uwe,Philcox:2023psd} and compression onto a one-dimensional skew-spectrum \citep{Munshi:2009ik}. Although such approaches have significant utility in constraining late-Universe signatures \citep[e.g.,][]{Coulton:2019bnz}, they are difficult to use as a generic primordial physics probe, since one must perform a costly set of integrals to convert from primordial- to CMB-space \citep[e.g.,][]{Philcox:2024wqx}. A powerful alternative is to perform a basis expansion in \textit{primordial} space, for example via the modal and \textsc{cmb-best} formalisms \citep{2009PhRvD..80d3510F,Fergusson:2009nv,Sohn:2023fte}. By expressing $B_\zeta$ as a polynomial series with coefficients computed from the CMB data, these methods can be used to cheaply constrain arbitrary inflationary models, and have recently used to constrain a variety of particle-exchange scenarios \citep{Sohn:2024xzd,Suman:2025tpv,Suman:2025vuf,Kumar:2026ogn,Salcedo:2026sdn,Cassem:2026ygh}. 

In this \textit{Letter}, we go one step further by exploiting the (assumed) dilatation symmetry of inflation. In particular, we directly estimate the physical quantity predicted by inflationary models: the scale-invariant bispectrum shape, $S^{\rm 2D}(x,y)$.\footnote{\resub{Available at \href{https://github.com/oliverphilcox/Binned-Bispectrum-Shape}{github.com/oliverphilcox/Binned-Bispectrum-Shape}.}} This approach, which practically corresponds to reconstructing the underlying shape in a set of logarithmic bins, has a number of practical benefits. Firstly, we can directly visualize the two-dimensional shape and its squeezed limit (which is home to the collider oscillations); in addition to observational interrogation, this allows one to assess the constraining power of different triangle configurations. Secondly, we can trivially compare theory and data for any scale-invariant model of interest (usually in a few milliseconds). Thirdly, since we work in logarithmic space and restrict to scale-invariant configurations, our basis is efficient and quasi-optimal for realistic inflationary models, including those with fast oscillations in the squeezed limit. Finally, we can build analogous \textit{scale-dependent} estimators, which allow the data to be interrogated in new ways (for example, assessing the information content as a function of $k$). To showcase our approach, we apply it to \textit{Planck} data, using the resulting measurements to constrain inflationary particle-exchange using more than $20\,000$ models of varying mass, spin, and sound-speed, computed using bootstrap techniques \citep{Arkani-Hamed:2018kmz,Pimentel:2022fsc,Jazayeri:2022kjy}. This opens the door to a wide variety of inflationary studies, which are limited only by the speed of computing theoretical models.

\section*{Estimators}
\noindent Novel processes occurring in the early Universe source non-Gaussianities in the distribution of the gauge-invariant curvature perturbation, $\zeta$. At leading-order, this is described by the two- and three-point functions
\beq
    \av{\zeta(\vk_1)\zeta(\vk_2)} &=& \delD{\vk_1+\vk_2}P_\zeta(k_1)\\\nonumber
    \av{\zeta(\vk_1)\zeta(\vk_2)\zeta(\vk_3)} &=& \delD{\vk_1+\vk_2+\vk_3}\\\nonumber
    &&\,\times\,B_\zeta(k_1,k_2,k_3),
\eeq
positing translation and rotation invariance. Assuming scale-invariance (with $n_s=1$), the power spectrum and bispectrum scale as $P_\zeta(k) \sim k^{-3}$ and $B_\zeta(k,k,k)\sim k^{-6}$ \citep[e.g.,][]{2020A&A...641A...6P}, which motivates defining the three-dimensional shape function, $S^{\rm 3D}$:
\beq
    B_\zeta(k_1,k_2,k_3) &\equiv& \frac{18}{5}\left[P_\zeta(k_1)P_\zeta(k_2)P_\zeta(k_3)\right]^{2/3}\\\nonumber
    &&\,\times\,S^{\rm 3D}(k_1,k_2,k_3),
\eeq
which is independent of $k$ if the bispectrum scales as $k^{2(n_s-4)}$ (matching that of $P^2_\zeta$). For convenience, we have absorbed the amplitude parameter into the shape function, such that $S^{\rm 3D}(k_{\rm pivot},k_{\rm pivot},k_{\rm pivot})=f_{\rm NL}$. 

In this \textit{Letter}, we seek to reconstruct the shape function from CMB observations. To this end, we expand $S^{\rm 3D}$ in a set of $k$-bins: 
\beq\label{eq: binned-shape}
    S^{\rm 3D}(k_1,k_2,k_3) &\approx & \sum_{n=1}^{N_{\rm bin}^{\rm 3D}}\frac{s^{\rm 3D}_n}{\Delta_n}\left[\Theta(k_1;\bar{k}_{n,1})\Theta(k_2;\bar{k}_{n,2})\right.\\\nonumber
    &&\qquad\qquad\,\times\,\left.\Theta(k_3;\bar{k}_{n,3})+\text{5 perms.}\right],
\eeq
where $\Theta$ is unity if $k$ is in the bin with center $\bar{k}$ and zero else, and $\Delta_n$ is a degeneracy factor, equal to six, two, or one, for equilateral, isosceles and scalene triangles respectively.\footnote{To enforce the triangle conditions and remove degeneracies, we set $|\bar{k}_{n,1}-\bar{k}_{n,2}|\leq \bar{k}_{n,3}\leq \bar{k}_{n,1}+\bar{k}_{n,2}$, and fix $\bar{k}_{n,1}\leq \bar{k}_{n,2}\leq \bar{k}_{n,3}$.} This compresses the three-dimensional bispectrum into a set of $N_{\rm bin}^{\rm 3D}$ coefficients $\{s^{\rm 3D}_n\}$. In practice, we will use a logarithmic binning in $k$-space to define \eqref{eq: binned-shape}; this efficiently captures information from the wide range of scales accessible to CMB datasets and retains sensitivity to highly squeezed configurations.

Since the CMB is a linear tracer of the primordial curvature perturbation, we can estimate $s^{\rm 3D}_n$ by correlating the temperature (or polarization) anisotropies at three points in space. As shown in previous works \citep[e.g.,][]{Philcox:2023psd,2011MNRAS.417....2S,2011arXiv1105.2791F,Komatsu:2003iq}, the minimum-variance estimator is given by
\beq\label{eq: 3d-estimator-main-text}
    \widehat{s}_{n}^{\rm 3D}&\sim &\sum_{\ell_im_i}\frac{\partial\av{a_{\ell_1m_1}a_{\ell_2m_2}a_{\ell_3m_3}}}{\partial s_n^{\rm 3D}}\big[\tilde{a}_{\ell_1m_1}\tilde{a}_{\ell_2m_2}\tilde{a}_{\ell_3m_3}\\\nonumber
    &&\qquad\qquad\qquad\qquad\qquad\,-3\tilde{a}_{\ell_1m_1}\av{\tilde{a}_{\ell_2m_2}\tilde{a}_{\ell_3m_3}}\big]
\eeq
(omitting the normalization factor), where $\tilde{a}_{\ell m}\equiv [C^{-1}a]_{\ell m}$ is an inverse-variance-weighted CMB map. As shown in the Supplemental Material, the estimator can be efficiently evaluated using
a combination of spherical harmonic transforms, Monte Carlo summation, and low-dimensional integration \citep[cf.,][]{2011MNRAS.417....2S,Philcox4pt1}; this is possible since \eqref{eq: binned-shape} is factorizable in momentum.

Under the assumption of scale-invariance, the physical quantity of interest is the \textit{two-dimensional} shape, $S^{\rm 2D}(x=k_1/k_3,y=k_2/k_3)$, rather than $S^{\rm 3D}(k_1,k_2,k_3)$. Building a direct estimator for $S^{\rm 2D}$ is computationally infeasible since the analog of \eqref{eq: 3d-estimator-main-text} is non-factorizable in momentum; however, one can proceed indirectly by judiciously combining estimates of the three-dimensional shape:
\beq\label{eq: 3d-to-2d-main-text}
    \widehat{S}^{\rm 2D}(x,y) &=& \frac{\sum_{k}\widehat{S}^{\rm 3D}(kx,ky,k)/\mathrm{var}[\widehat{S}^{\rm 3D}(kx,ky,kz)]}{\sum_{k}1/\mathrm{var}[\widehat{S}^{\rm 3D}(kx,ky,kz)]},
\eeq
noting that $S^{\rm 3D}(kx,ky,k)=S^{\rm 2D}(x,y)$ in the de Sitter limit.
As discussed in the Supplemental Material, this can be rewritten as an estimator for a two-dimensional shape coefficient, $s_N^{\rm 2D}$, encoding the amplitude of the shape in a bin with side ratios $\bar{x}_N, \bar{y}_N$ (with $0\leq \bar{x}_N\leq \bar{y}_N\leq 1$ and $|\bar{x}_N-\bar{y}_N|\leq 1$). Due to our choice of logarithmic $k$-space binning, the sum over $k$ can be absorbed within the estimator, such that the computation time does not scale with the number of three-dimensional bins. The net result is an efficient estimator for the binned two-dimensional shape that can be evaluated using spherical harmonic transforms.

The above estimator has been added to the \textsc{PolySpec} package \citep{PolyBin,Philcox4pt2}.\footnote{Available at \href{https://github.com/oliverphilcox/PolySpec}{GitHub.com/OliverPhilcox/PolySpec}.} Starting from an input set of temperature and polarization maps, the code estimates the binned bispectrum coefficients (and the corresponding Fisher matrices), given a set of $k$-bin edges and a scheme for combining them into bins. The measurements are unbiased by the experimental mask and beam, explicitly subtract foreground contributions sourced by ISW-lensing cross-correlations \citep{Hill:2018ypf,Philcox:2025lxt}, and are close to minimum variance.

\section*{The Planck Shape Function}

\begin{figure}
    \centering
    \includegraphics[width=0.85\linewidth]{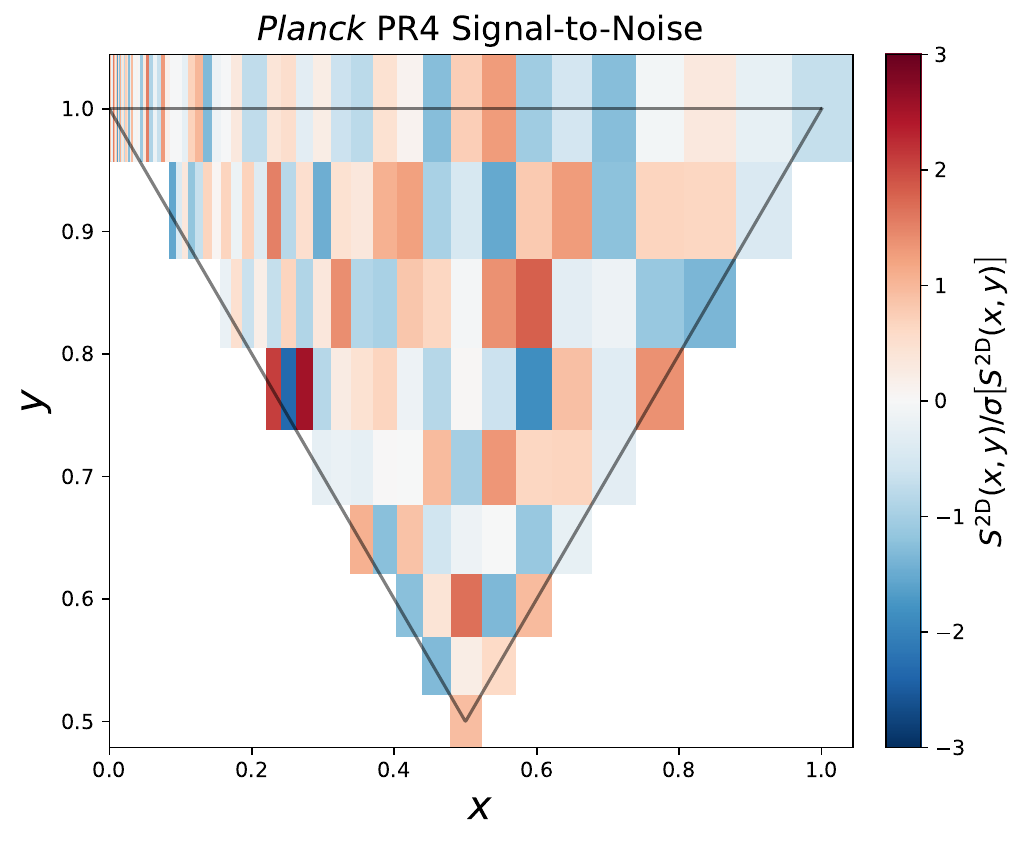}
    \includegraphics[width=0.85\linewidth]{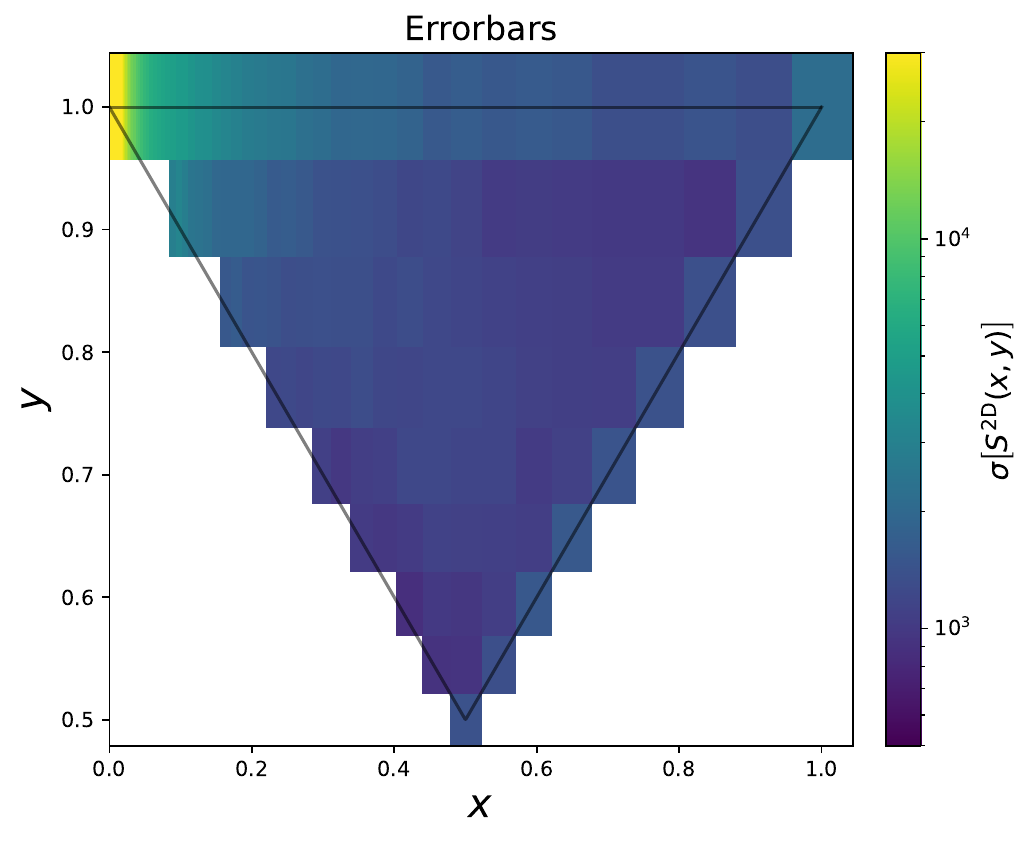}
    \caption{Two-dimensional shape functions, $S^{\rm 2D}(x,y)$, reconstructed from \textit{Planck} PR4 temperature and polarization anisotropies. The top panel shows the signal-to-noise ratio in each $x,y$-bin (imposing the triangle conditions, with $x\leq y\leq 1$), whilst the bottom shows the empirical error, obtained from $400$ FFP10 simulations and normalized such that $S^{\rm 2D}(1,1)=f_{\rm NL}$. The constraints have similar magnitude across most of the $(x,y)$-plane, but are significantly degraded in the squeezed limit. Visually, we find no evidence for a primordial signal with $\chi^2=175.9$ for $171$ degrees of freedom. By comparing these measurements with theoretical shape functions, we can immediately place constraints on a wealth of scale-invariant models; an example is shown in Fig.\,\ref{fig: bootstrap-constraints}.}
    \label{fig: shape-functions-2d}
\end{figure}

\begin{figure}
    \centering
    \includegraphics[width=0.9\linewidth]{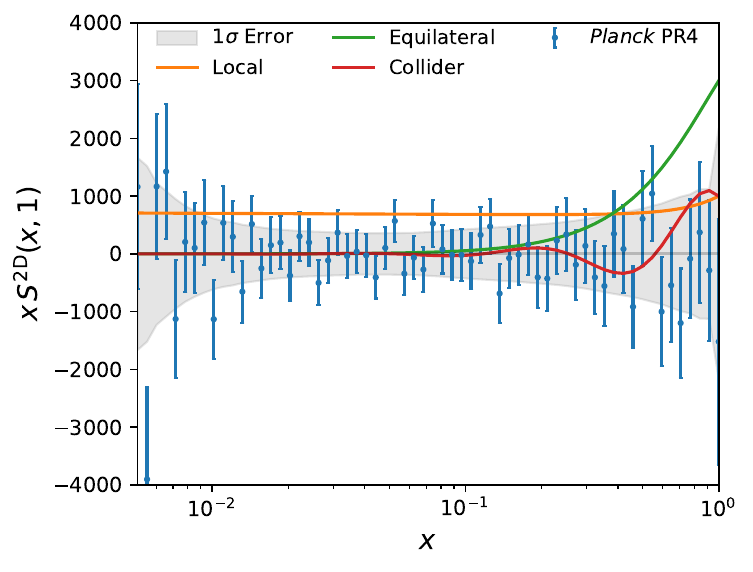}
    \caption{Squeezed limit of the primordial shape function reconstructed from \textit{Planck} temperature and polarization data, as a function of $x$ ($=k_{\rm long}/k_{\rm short}$). Shaded regions and errorbars show the theoretical and empirical errors from the inverse Fisher matrix and FFP10 simulations respectively, and we omit configurations with $x<0.005$, which cannot be meaningfully constrained using \textit{Planck} data. For reference, we show the scalings of three fiducial models (with arbitrary normalization): local (orange), equilateral (green), and the weakly-mixed cosmological collider (red; with $m\approx 5H$). Notably, the shape noise scales as $1/x$; this makes it difficult to probe the collider oscillations ($\propto x^{1/2}$) in practice.}
    \label{fig: squeezed}
\end{figure}

\noindent Using the above technology, we can reconstruct the two-dimensional shape function from a set of CMB observations. Here, we apply this to temperature and polarization maps from the \textit{Planck} PR4 release \citep{Rosenberg:2022sdy,Planck:2020olo,Tristram:2020wbi}, computing the bin amplitudes across $171$ $(x,y)$-pairs (themselves assembled from $5424$ $(k_1,k_2,k_3)$-configurations) down to $x_{\rm min}\approx 10^{-3}$. The shape measurements are displayed in Fig.\,\ref{fig: shape-functions-2d}, with the squeezed limit shown in Fig.\,\ref{fig: squeezed}. In the Supplemental Material we give further details on the analysis methods and present additional results including a study of the scale-dependence of our inflationary constraints.

From a visual inspection, the two-dimensional shape-function shows no evidence for new inflationary physics, with a combined $\chi^2$-per-degree of freedom of $1.03$, and a maximum signal-to-noise of $2.5\sigma$.\footnote{Accounting for look-elsewhere effects, this is consistent with the null hypothesis at $0.9\sigma$. We note that neighboring points in the $(x,y)$-plane are correlated.} The errors on $S^{\rm 2D}(x,y)$ have a similar magnitude across much of the two-dimensional space, with $\sigma\sim 10^3$ for all $x>0.05$ (in units where $S^{\rm 2D}(1,1)=f_{\rm NL}$); the uncertainty grows dramatically in the squeezed limit, due to the larger cosmic variance associated with soft modes \citep{Kalaja:2020mkq}. In the Supplemental Material, we demonstrate that the errorbars are consistent with the theoretical expectation, \textit{i.e.}\ that our estimator is close to optimal.

In Fig.\,\ref{fig: squeezed}, we plot the squeezed limit of the primordial bispectrum (which is the quantity considered in many theoretical studies), finding no evidence for new physics. In the moderately squeezed regime, the errors scale as $x^{-1}$; for $x\lesssim 10^{-2}$, the errors inflate dramatically, indicating that highly squeezed configurations are difficult to constrain using current CMB data (though this will improve with next-generation datasets, such as the Simons Observatory). Notably, the characteristic oscillations indicative of massive particle exchange scale as $x^{1/2}$ in the squeezed limit \citep[e.g.,][]{Arkani-Hamed:2015bza}; comparing this to the empirical scaling of $\sigma[S^{\rm 2D}]$, we we conclude that the CMB is only able to probe the cosmological collider paradigm on intermediate scales.

\section*{Model Constraints}
\noindent By combining our measurements of the dimensionless shape function with theoretical predictions, we can straightforwardly obtain bounds on a wide variety of scale-invariant inflationary models. In the Supplemental Material, we demonstrate that our approach yields constraints on factorizable shapes that match those from optimal template-specific estimators within $\approx 10\%$; here, we demonstrate the approach by constraining non-factorizable models drawn from the cosmological collider program.

Specifically, we search for bispectra induced by the exchange of weakly-mixed boost-breaking spin-zero, spin-one and spin-two particles, across a spectrum of mass parameters, $\mu$ (\resub{from sub-conformal masses} up to six times the inflationary Hubble scale), and relative sound-speeds, $c_s$.\footnote{Due to the availability of bootstrap calculations, we limit ourselves to single-exchange interactions in this \textit{Letter}. As discussed in \citep[e.g.,][]{Chen:2009we,Chen:2017ryl,Kumar:2026ogn,cosmoflow1,cosmoflow2}, larger spin-zero signals can be generated from double- and triple-exchange diagrams.} 
In contrast to previous CMB studies \citep{Sohn:2024xzd,Suman:2025vuf,Suman:2025tpv}, we use the full shape function templates computed using the cosmological bootstrap \citep{Pimentel:2022fsc} (\resub{in place of analytic approximations}), extending the calculation to $c_s<1$ following \citep{Cabass:2024wob} and paying close attention to numerical convergence. For each model, we constrain an overall amplitude $f_{\rm NL}$ (set by the couplings in the effective Lagrangian), encoding the size of the signal in the equilateral limit. In full, we evaluate the shape functions in $171$ triangle configurations, across more than $6000$ combinations of mass and spin and four types of interaction. Whilst the theoretical computation is fairly expensive (requiring $\approx 100$ hours in \textsc{mathematica}), the application to data is not: we obtain all of the $f_{\rm NL}$ measurements and errorbars in just $0.6\,\mathrm{s}$.

\begin{figure*}
    \centering
    \includegraphics[width=0.84\linewidth]{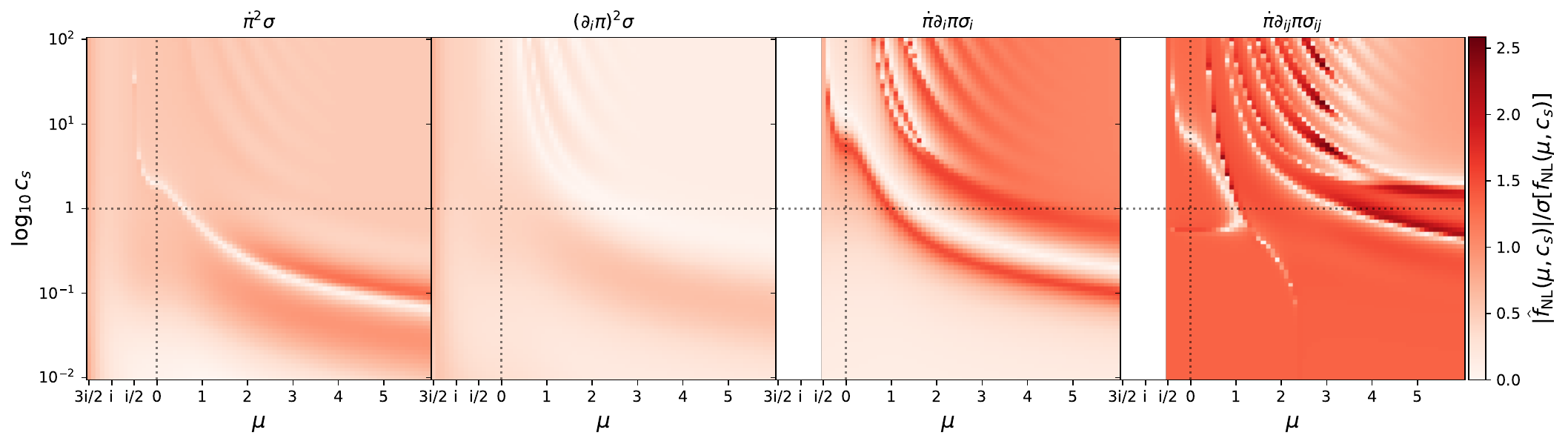}
    \includegraphics[width=0.84\linewidth]{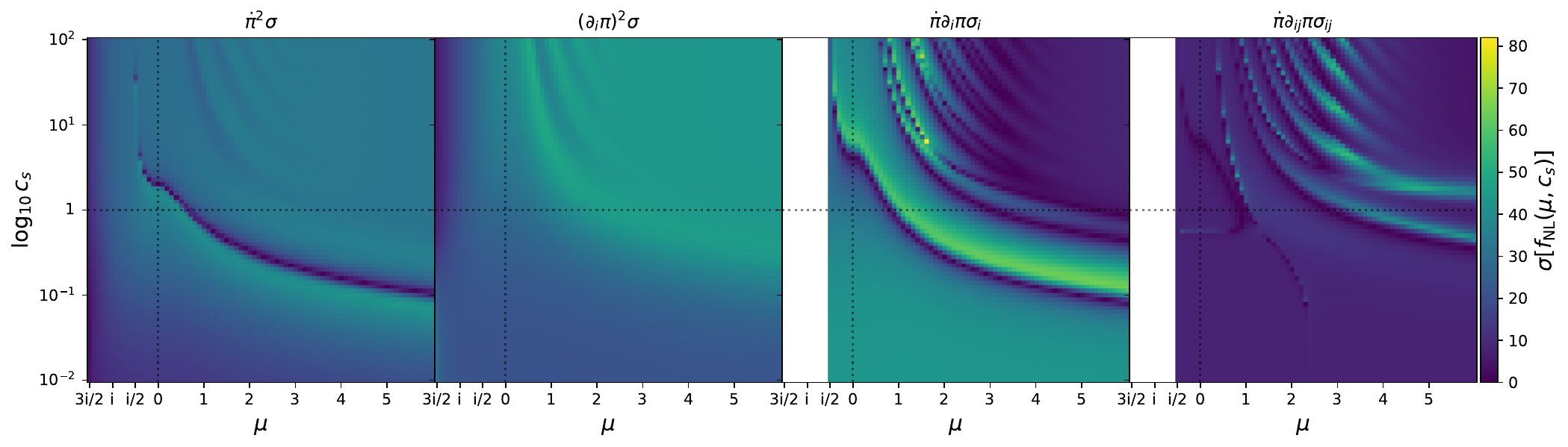}
    \caption{Constraints on the exchange of massive particles in inflation. We show results for four choices of interaction (induced by the cubic vertices shown in the titles; see the Supplemental Material for the underlying Lagrangian), as a function of mass parameter $\mu\equiv \sqrt{m^2_\sigma/H^2-9/4}$ (or $\sqrt{m^2_\sigma/H^2-(s-1/2)^2}$ for spin $s>0$), relative sound-speed $c_s=c_\pi/c_\sigma$, and spin (zero, zero, one, and two respectively). For higher-spin exchange, we restrict to $\mathrm{Im}[\mu]\leq 1/2$ according to the Higuchi bound \citep{Higuchi:1986py}. The theoretical models are computed numerically using bootstrap techniques (in contrast to most previous works, which used \resub{analytic approximations}). The top panel shows the signal-to-noise ratio, whilst the bottom gives the $1\sigma$ error on the $f_{\rm NL}$ amplitude. We report no detection of collider non-Gaussianity with a maximum detection significance of $2.6\sigma$ (for spin-two with $\mu \approx 3, c_s\approx 5$), and place strong constraints on the models across a broad range of masses and spins. Given the shape function measurements from Fig.\,\ref{fig: shape-functions-2d}, computation of these constraints required just $0.6$ seconds.}
    \label{fig: bootstrap-constraints}
\end{figure*}

Fig.\,\ref{fig: bootstrap-constraints} shows the constraints on the template amplitudes. 
Across all models, we find a maximum detection significance of $2.6\sigma$ (from the spin-two template), indicating no evidence for new physics. The complex structure in the $(\mu,c_s)$-plane arises due to our choice of normalization (with ridges when the shape function crosses zero) and the complex interplay between oscillations and smooth components in the underlying templates. For $c_s\gtrsim 1$ and large $\mu$, the exchange shapes are dominated by a smooth contribution indistinguishable from single-field self-interactions; more novel phenomenology arises for low sound-speeds and masses closer to the Hubble scale. The $f_{\rm NL}$ bounds from Fig.\,\ref{fig: bootstrap-constraints} can be translated into constraints on the inflationary microphysics; this is explored in \citep{cosmoflow1}. 

\section*{Discussion}
\noindent The above results elucidate an exciting opportunity for inflationary cosmology: using the reconstructed shape functions, we can quickly and accurate place constraints on \textit{any} scale-invariant inflationary model. In this \textit{Letter}, we have considered a single class of models, but there are many others worthy of exploration, including inflationary self-interactions \citep[e.g.,][]{Chen:2006nt,Chen:2008wn,Cheung:2007st,Senatore:2009gt}, multiple exchanges \citep[e.g.,][]{Lee:2016vti,Xianyu:2023ytd}, strongly-mixed sectors \citep[e.g.,][]{Werth:2023pfl,Pinol:2023oux}, strongly-coupled sectors \citep[e.g.,][]{An:2017hlx,Pimentel:2025rds,Jiang:2025mlm,Green:2013rd}, chemical potentials \citep[e.g.,][]{Chen:2018xck,Wang:2019gbi,Bodas:2020yho,Sou:2021juh,Tong:2022cdz}, loop diagrams \citep[e.g.,][]{Senatore:2009cf,Xianyu:2022jwk,Qin:2022lva,Bodas:2025vpb}, and dissipative effects \citep[e.g.,][]{Salcedo:2024smn}. Since we do not require that the input templates are factorizable, we can directly analyze numerically-computed bispectra, allowing connection with recent inflationary developments including the cosmological bootstrap \citep[e.g.,][]{Arkani-Hamed:2018kmz,Pajer:2020wnj}, Mellin representations \citep[e.g.,][]{Qin:2022lva,Qin:2022fbv,Sleight:2019hfp}, and flow-based methods \citep[e.g.,][]{Pinol:2023oux}. Finally, one can analyze families of models simultaneously; this could facilitate a direct analysis of the effective inflationary Lagrangian, accounting for non-linearly realized symmetries. 

An attractive feature of our binned bispectrum decomposition is that it allows distinct experiments to be compared in the same space. In particular, it would be interesting to apply this to high-resolution data from the Simons Observatory \citep{SimonsObservatory:2018koc}; in addition to reducing the errorbars on all scales (due to larger $\ell_{\rm max}$), this would significantly enhance constraints on highly-squeezed configurations. Moreover, one could imagine performing a similar reconstruction using large-scale structure data; though this is complicated by non-linear structure formation, it would give new insights on the inflationary landscape and allow for straightforward combination of CMB and galaxy survey datasets.

Finally, the methodology from this \textit{Letter} could be developed in new directions. A straightforward extension would be to the \textit{tensor} shape function \citep[cf.,][]{Duivenvoorden:2019ses,Philcox:2024wqx,Philcox:2023xxk}. This would greatly simplify analyses of various higher-spin inflationary models and could be achieved by adding polarization factors and exchanging scalar transfer functions for their tensor equivalents. Even more excitingly, one could build a binned estimator for the scale-invariant inflationary trispectrum (building on the modal decompositions of \citep{Fergusson:2010gn}). Since the four-point function is a five-dimensional object, this is highly non-trivial, but could yield important constraints on inflationary scattering processes. We leave these extensions (and many others) to the future.


\vskip 8pt
\acknowledgments
{\footnotesize
\begingroup
\hypersetup{hidelinks}
\noindent 
We thank Sam Goldstein, Sadra Jazayeri, Gui Pimentel, Xi Tong, Dong-Gang Wang, Denis Werth, Zhong-Zhi Xianyu, and Chen Yang for insightful discussions and comments on the draft. OHEP acknowledges support from the \href{https://www.flickr.com/photos/198816819@N07/55151059948/}{SWISS airllines fllight crew}. This \textit{Letter} was inspired by conversations at the 41st Annual Colloquium of the Institut d'Astrophysique de Paris. All computations were run at facilities supported by the Scientific Computing Core at the Flatiron Institute, a division of the Simons Foundation. 
\endgroup
\vskip 4pt
}


\clearpage
\onecolumngrid

\setcounter{equation}{0}
\setcounter{figure}{0}
\setcounter{table}{0}
\setcounter{section}{0}

\renewcommand{\theequation}{S\arabic{equation}}
\renewcommand{\thefigure}{S\arabic{figure}}
\renewcommand{\thetable}{S\arabic{table}}

\begin{center}
{\large \textbf{Supplemental Material}}
\end{center}






\section{Binned Shape Function Estimators}
\subsubsection{Three Dimensions}
\noindent Below, we outline the derivation and practical computation of the binned shape function estimators discussed in the main text. Given a harmonic-space CMB map $a_{\ell m}^{{\rm obs},X}$ (where $X$ indexes the field, \textit{i.e.}\ $X\in\{T,E\}$, noting that $B$-modes do not trace scalar perturbations), a general estimator for a set of quantities $\{s^{\rm 3D}_n\}$ appearing in the CMB bispectrum is given by $\widehat{s}_{n}^{\rm 3D}\equiv [\F^{-1}\widehat{N}]^{\rm 3D}_n$, where
\beq\label{eq: 3d-estimator}
    \widehat{N}^{\rm 3D}_n &=& \frac{1}{6}\sum_{\ell_im_iX_i}\frac{\partial \av{a_{\ell_1m_1}^{X_1}a_{\ell_2m_2}^{X_2}a_{\ell_3m_3}^{X_3}}}{\partial s^{\rm 3D}_n}\left(\tilde{a}_{\ell_1m_1}^{X_1}\tilde{a}_{\ell_2m_2}^{X_2}\tilde{a}_{\ell_3m_3}^{X_3}-\left[\tilde{a}_{\ell_1m_1}^{X_1}\av{\tilde{a}_{\ell_2m_2}^{X_2}\tilde{a}_{\ell_3m_3}^{X_3}}+\text{2 perms.}\right]\right)^*\\\nonumber
    \F^{\rm 3D}_{nn'} &=& \frac{1}{6}\sum_{\ell_im_iX_i\ell'_im'_iX'_i}\frac{\partial \av{a_{\ell_1m_1}^{X_1}a_{\ell_2m_2}^{X_2}a_{\ell_3m_3}^{X_3}}}{\partial s^{\rm 3D}_n}\left([\Si\mathsf{P}]^{X_1X_1'}_{\ell_1\ell_1',m_1m_1'}[\Si\mathsf{P}]^{X_2X_2'}_{\ell_2\ell_2',m_2m_2'}[\Si\mathsf{P}]^{X_3X_3'}_{\ell_3\ell_3',m_3m_3'}\right)\left(\frac{\partial \av{a_{\ell_1'm_1'}^{X_1'}a_{\ell_2'm_2'}^{X_2'}a_{\ell_3'm_3'}^{X_3'}}}{\partial s^{\rm 3D}_{n'}}\right)^*
\eeq
\citep[e.g.,][]{Philcox4pt1,Philcox:2023psd,2011MNRAS.417....2S,2009PhRvD..80d3510F,2011arXiv1105.2791F}. Here, $\tilde{a}\equiv \Si\,a^{\rm obs}$ for some weighting operator $\Si$, and $\mathsf{P}$ is the pointing matrix, encoding the response of the observed field $a^{\rm obs}$ to the underlying CMB realization $a$: $a^{\rm obs}=\mathsf{P}a+\text{noise}$. This is naught but a matched filter applied to the $\Si$-weighted data, with the normalization fixed by requiring that the estimator is unbiased for arbitrary $\Si$. Setting $\Si\mathsf{P}$ equal to the inverse covariance matrix of $a^{\rm obs}$, the estimator satisfies its Cram\'er-Rao bound for $\{s_n^{\rm 3D}\}$, achieving minimum-variance. In this limit, the estimator has covariance $\mathcal{F}^{-1,\rm 3D}$, implying that $\F^{\rm 3D}$ is the (mask- and beam-dependent) Fisher matrix.

To evaluate the above expression, we require the theoretical bispectrum $\av{a_{\ell_1m_1}^{X_1}a_{\ell_2m_2}^{X_2}a_{\ell_3m_3}^{X_3}}$, This can be computed starting from the standard relation between the underlying CMB anisotropies and the curvature perturbations \citep[e.g.,][]{Komatsu:2001rj}
\beq   
    a_{\ell m}^X &=& 4\pi i^\ell\int\frac{d\vk}{(2\pi)^3}\mathcal{T}_\ell^X(k)\zeta(\vk)Y^*_{\ell m}(\hk).
\eeq
Using the definition of the binned primordial shape and simplifying, we find
\beq\label{eq: bin-derivative}
    \frac{\partial \av{a_{\ell_1m_1}^{X_1}a_{\ell_2m_2}^{X_2}a_{\ell_3m_3}^{X_3}}}{\partial s^{\rm 3D}_n} &=& \frac{18}{5}\frac{1}{\Delta_n}\int r^2dr\,d\hr\,\prod_{i=1}^3\left[f_{\ell_i}^{X_i}(r;\bar{k}_{n,i})Y^*_{\ell_im_i}(\hr)\right]\,+\text{5 perms.},
\eeq
where we have defined the filters
\beq
    f_{\ell}^{X}(r;\bar{k}) &\equiv& (-1)^{\ell}\frac{2}{\pi}\int k^2dk\,P_\zeta^{2/3}(k)\mathcal{T}^{X}_{\ell}(k)j_{\ell}(kr)\Theta(k;\bar{k}_n).
\eeq
This involves an integral over $r$ (and over $\hr$, which could be rewritten as a Gaunt symbol); this arises from the momentum-conserving Dirac delta. Inserting \eqref{eq: bin-derivative} into \eqref{eq: 3d-estimator}, we find a combined estimator, which is analogous to the canonical Komatsu-Spergel-Wandelt form (KSW; \citep{Komatsu:2003iq}):
\beq\label{eq: 3d-numerator-simplified}
    \widehat{N}^{\rm 3D}_n &=& \frac{18}{5}\frac{1}{\Delta_n}\int d\vr\,\bigg\{F_{n,1}[\tilde{a}](\vr)F_{n,2}[\tilde{a}](\vr)F_{n,3}[\tilde{a}](\vr)-\left(F_{n,1}[\tilde{a}](\vr)\av{F_{n,2}[\tilde{a}](\vr)F_{n,3}[\tilde{a}](\vr)}+\text{2 perms.}\right)\bigg\}
\eeq
defining
\beq
    F_{n,i}[\tilde{a}](\vr) &=& \sum_{\ell m X}Y_{\ell m}(\hr)f_{\ell}^X(r;\bar{k}_{n,i})\tilde{a}_{\ell m}^{X}.
\eeq
Expression \eqref{eq: 3d-numerator-simplified} can be efficiently evaluated using one-dimensional integration (for $\int k^2dk$ and $\int r^2dr$), spherical harmonic transforms (for $F$), and summation over pixels (for $\int d\hr$), with combined complexity $\mathcal{O}(N_{\rm pix}\log N_{\rm pix})$ for $N_{\rm pix}$ pixels. This is a significant improvement over the na\"ive estimator \eqref{eq: 3d-estimator}, which has complexity $\mathcal{O}(N_{\rm pix}^3)$. The linear term in $\widehat{N}^{\rm 3D}$ can be evaluated using a set of realistic simulations to evaluate the expectation, \textit{i.e.}\ setting $\av{F[\tilde{a}]F[\tilde{a}]}\to (1/N)\sum_{i=1}^N F[\tilde{\alpha}_i]F[\tilde{\alpha}_i]$ for a suite of $N$ simulations $\{\alpha_i\}$.

To compute the normalization matrix $\F$, we follow the approach of \citep{2011MNRAS.417....2S} (see also \citep{girard89,hutchinson90,Philcox4pt1}), splitting up the high-dimensional sum over $\ell_i,\ell_i'$ by introducing a set of independent and identically distributed random fields $\{\alpha^{(i)}\}$ for $i\in\{1,2\}$:
\beq
    \F^{\rm 3D}_{nn'} &=& \frac{1}{6}\sum_{\ell\ell'mm'XX'}\left\langle\left(Q^{X,\rm 3D}_{\ell m,n}[\mathsf{A}^{-1}\alpha^{(1)},\mathsf{A}^{-1}\alpha^{(2)}]\right)[\Si\mathsf{P}]^{XX'}_{\ell\ell',mm'}\left(Q^{X',\rm 3D}_{\ell'm',n'}[\mathsf{P}\Si \alpha^{(1)},\mathsf{P}\Si \alpha^{(2)}]\right)^*\right\rangle_{\{\alpha^{(1)},\alpha^{(2)}\}},
\eeq
where $\av{\alpha^{(i)}\alpha^{(j)}} = \delta^{ij}_{\rm K}\mathsf{A}$ for invertible $\mathsf{A}$. This involves the derivatives
\beq
    Q^{X, \rm 3D}_{\ell m, n}[x,y] &=& \sum_{\ell_2\ell_3m_2m_3X_2X_3}\frac{\partial \av{a_{\ell m}^{X}a_{\ell_2m_2}^{X_2}a_{\ell_3m_3}^{X_3}}}{\partial s^{\rm 3D}_n}x^{X_2*}_{\ell_2m_2}y^{X_3*}_{\ell_3m_3},
\eeq
which can be computed using analogous methods to above, leading to
\beq\label{eq: Q-deriv3D}
    Q^{X, \rm 3D}_{\ell m, n}[x,y] &=& \frac{18}{5}\frac{1}{\Delta_n}\int r^2dr\,f_{\ell}^{X}(r;\bar{k}_{n,1})\int d\hr\,Y^*_{\ell m}(\hr)F_{n,2}[x](\vr)F_{n,3}[y](\vr)\,+\text{5 perms.},
\eeq
with permutations over the bin orderings. Practically, computation of $\F^{\rm 3D}$ reduces to a set of harmonic transforms and low-dimensional integrals, following a sum over a small set of Monte Carlo maps $\{\alpha^{(i)}\}$. This form of $\F$ fully accounts for the observational mask, beam, and noise properties, going beyond the simplified diagonal limits used in most CMB estimators \citep[e.g.,][]{Sohn:2023fte,Fergusson:2009nv}.

\subsubsection{Two Dimensions}
\noindent As motivated in the main text, we can form an estimator for the two-dimensional shape $S^{\rm 2D}$ by averaging over three-dimensional measurements $\{\widehat{s}^{\rm 3D}_n\equiv \widehat{S}^{\rm 3D}(k_{n,1},k_{n,2},k_{n,3})\}$ with a given set of side ratios. Defining the two-dimensional bispectrum coefficients $\{s^{\rm 2D}_N\}\equiv \{S^{\rm 2D}(\bar{x}_N,\bar{y}_N)\}$, the best estimator for $s^{\rm 2D}$ from $s^{\rm 3D}$ is given by:\footnote{Since we have performed a pre-compression into $\{s^{\rm 3D}\}$, this is not strictly the optimal estimator for $s^{\rm 2D}$ from the observed CMB fluctuations. The fully optimal estimator is not factorizable, and thus cannot be practically implemented; however, \eqref{eq: 3d-to-2d} provides an excellent approximation, which becomes exact in the limit of thin $k$-bins.}
\beq\label{eq: 3d-to-2d}
    \widehat{s}^{\rm 2D}_N &\propto & \sum_{n\in N}\left[\mathrm{Cov}^{-1}_{\rm 3D}\cdot\widehat{s}^{\rm 3D}\right]_{n},
\eeq
dropping a normalization term and summing only over bins satisfying $\bar{x}_N = \bar{k}_{n,1}/\bar{k}_{n,3}$, $\bar{y}_N = \bar{k}_{n,2}/\bar{k}_{n,3}$, where $\mathrm{Cov}_{\rm 3D}$ is the covariance matrix of the three-dimensional estimator,. Inserting \eqref{eq: 3d-estimator} and noting that $\mathrm{Cov}^{-1}_{\rm 3D}$ is equal to $\F^{\rm 3D}$ (in the optimal limit), we obtain a practical estimator for the two-dimensional coefficients:
\beq\label{eq: 2d-estimator}
    &&\widehat{s}^{\rm 2D}_N \equiv [\F^{-1}\widehat{N}]^{\rm 2D}_N, \qquad \widehat{N}^{\rm 2D}_{N} = \sum_n \P_{Nn}\widehat{N}^{\rm 3D}_n, \qquad \F^{\rm 2D}_{NN'} = \sum_{nn'}\P_{Nn}\F^{\rm 3D}_{nn'}\P_{N'n'},
\eeq
where the projection matrix $\P_{Nn}$ transforms from three-dimensional triangle bins $\{\bar{k}_{n,1},\bar{k}_{n,2},\bar{k}_{n,3}\}$ to two-dimensional shape bins $\{\bar{x}_N,\bar{y}_N\}$, with $\bar{x}_N = \bar{k}_{n,1}/\bar{k}_{n,3}$, $\bar{y}=\bar{k}_{n,2}/\bar{k}_{n,3}$, as before. 

Since estimator \eqref{eq: 2d-estimator} is a linear transformation of the three-dimensional estimator, it can be similarly written in terms of spherical harmonic transforms, low-dimensional integrals, and Monte Carlo sums (and thus efficiently computed). When assembling the numerator, $\widehat{N}^{\rm 2D}$, the rate-limiting step is computation of the filtered fields, $F_{n,i}$, which scales with the number of one-dimensional $k$-bins $\bar{k}_{n,i}$ rather than the number of three-dimensional bin triplets.\footnote{This assumes that each one-dimensional bin contributes to multiple three-dimensional configurations, which is the case for the logarithmic binning used in this \textit{Letter}.} When computing the normalization matrix $\F^{\rm 2D}$, we require the following derivative:
\beq
    Q^{X, \rm 2D}_{\ell m, N}[x,y] &=& \frac{18}{5}\sum_n\frac{\P_{Nn}}{\Delta_n}\int r^2dr\,f_{\ell}^{X}(r;\bar{k}_{n,1})\int d\hr\,Y^*_{\ell m}(\hr)F_{n,2}[x](\vr)F_{n,3}[y](\vr)\,+\text{5 perms.}.
\eeq
Computation of the $Q$ derivatives can be organized in such a way to scale with the number of unique choices of $\bar{k}_{n,2}$ and $\bar{k}_{n,3}$ (which scales with the number of $x,y$-bins), rather than the total number of bin triplets. The net result is that we can compute the two-dimensional estimator from a large number of bin triplets without undue computational expense.

\subsubsection{Binning}
\noindent To implement the estimators, we must specify the one-dimensional $k$-bins, which inform the allowed two-dimensional shape configurations. As motivated in the main text, we advocate for logarithmically-spaced $k$-bins with centers
\beq\label{eq: binning}
    \bar{k}\in \big\{k_{\rm min}, rk_{\rm min},r^2k_{\rm min},\cdots,k_{\rm max}\big\}
\eeq
for some $r>1$, where $k_{\rm max}=r^{n_{\rm bin}-1}k_{\rm min}$. With this definition, the associated two-dimensional bin-centers also form a geometric series:
\beq
    \bar{x} \in \{r^{1-n_{\rm bin}},r^{2-n_{\rm bin}}, \cdots, 1\}, \quad \bar{y}\in\{r^{-K},r^{1-K}, \cdots, 1\}
\eeq
(given the triangle conditions), where $K$ is the largest integer such that $r^{-K} \geq 1/2$. To ensure that the flattened configuration $x=y=1/2$ is contained within our dataset, we set $r = 2^{1/K}$ for integer $K\geq 1$. This approach is maximally efficient, with each two-dimensional bin being populated by a similar number of three-dimensional configurations.

Next, we define the expectation of the three- and two-dimensional estimators in terms of the true underlying shape functions $S^{\rm 3D}$ and $S^{\rm 2D}$. In the first case, we find
\beq\label{eq: S-3d-average}
    \mathbb{E}\left[\widehat{S}_{\rm 3D}(\bar{k}_1,\bar{k}_2,\bar{k}_3)\right] &\propto& \int_{\bar{k}_1/\sqrt{r}}^{\bar{k}_1\sqrt{r}} dk_1\int_{\bar{k}_2/\sqrt{r}}^{\bar{k}_2\sqrt{r}}dk_2\int_{\mathrm{max}(|k_1-k_2|,k_3/\sqrt{r})}^{\mathrm{min}(k_1+k_2,k_3\sqrt{r})}dk_3\,S^{\rm 3D}(k_1,k_2,k_3)
\eeq
(dropping a normalization factor), where we have ignored the variations of the transfer functions and experimental noise across a bin (noting that the bin-averaged effects are already accounted for in the global normalization, $\F^{\rm 3D}$). Assuming scale-invariance we can form the expectation of the two-dimensional estimator directly from \eqref{eq: S-3d-average}:
\beq\label{eq: S-2d-average}
    \mathbb{E}\left[\widehat{S}_{\rm 2D}(\bar{x},\bar{y})\right] &\propto& \int_{\bar{x}/\sqrt{r}}^{\bar{x}\sqrt{r}} dX\int_{\bar{y}/\sqrt{r}}^{\bar{y}\sqrt{r}}dY\int_{\mathrm{max}(|X-Y|,1/\sqrt{r})}^{\mathrm{min}(X+Y,\sqrt{r})}dR\,S^{\rm 2D}(X/R,Y/R).
\eeq
This contains contributions from triangles with $x\in[\bar{x}/r,\bar{x}r), y\in[\bar{y}/r,\bar{y}r)$ implying that (a) two-dimensional bins have twice the (logarithmic) width of three-dimensional bins, and (b) neighboring bins will be correlated. If the $k$-bins are sufficiently wide, it may be necessary to bin-integrate the input $S^{\rm 2D}$ model to compare theory and data; as shown below, this is not necessary for the scenarios considered herein.

\section{Application to Planck}
\subsubsection{Setup}
\noindent The estimators described above are now included in the \textsc{PolySpec} code \citep{PolyBin,Philcox4pt2}, which facilitates the quasi-optimal estimation of a range of bispectrum (and trispectrum) signatures, including binned estimators \citep{Philcox:2023uwe,Philcox:2023psd}, standard inflationary templates \citep{Philcox4pt2}, generalized neural-network basis functions \citep{Philcox:2025bbo}, and secondary effects, such as the ISW-lensing cross-correlation \citep{Philcox:2025lxt}. The code computes three quantities: the data-dependent numerator, $\widehat{N}$, the data-independent normalization, $\F$ (which accounts for the correlation between all shapes included), and the idealized normalization, $\F^{\rm ideal}$. The latter quantity, which is equal to $\F$ in the limit of translation-invariant noise and mask, is used to define the radial integration grid entering \eqref{eq: 3d-numerator-simplified}\,\&\,\eqref{eq: Q-deriv3D}. This follows the approach of \citep{2011MNRAS.417....2S}, optimizing for a set of radial points and weights that preserve the structure and amplitude of $\F^{\rm ideal}$ to a given tolerance. In principle, one should perform this optimization $\F^{\rm ideal}$ for every three-dimensional bin in the estimator; here, we adopt a simpler approach, optimizing just a single bin with $k_i\in[k_{\rm min},k_{\rm max}]$. This is found to be an excellent approximation in practice, with the errorbars changing by $<2\%$ if we instead optimize based on the $\F^{\rm ideal}$ matrix for the equilateral template.

Our dataset comprises the \textit{Planck} Public Release 4 (PR4) temperature and $E$-mode polarization anisotropies, processed with the \textsc{npipe} pipeline \citep{Planck:2020olo,Rosenberg:2022sdy,Tristram:2020wbi}. To remove foreground contamination, we utilize the \textsc{sevem} component-separation method \citep{Planck:2018yye} and apply the common component-separation masks in both temperature and polarization (inpainted using a linear scheme across $100$ iterations). Following previous works \citep{Philcox4pt3}, we deconvolve the \textit{Planck} beam, including the \textsc{npipe} low-$\ell$ transfer function, and apply an inverse-variance filtering, which involves the noise power spectrum measured from half-mission maps. $400$ FFP10/\textsc{npipe} simulations are used to compute covariances, with a further $N_{\rm lin}=50$ used to form the linear term for estimator \eqref{eq: 3d-estimator} \citep{Planck:2015txa}. To suppress the scatter induced by the finite $N_{\rm lin}$, we set $N_{\rm lin}=500$ when analyzing the observational data (noting that the restriction to $N_{\rm lin}=50$ in the simulations inflates the errorbars by $\lesssim 3\%$. Transfer functions are computed using the \textit{Planck} 2018 fiducial cosmology \citep{2020A&A...641A...6P}, working at high-resolution in both $k$ and $\ell$. Throughout, we assume an inflationary sector specified by an amplitude $A_s=2.10\times 10^{-9}$, a slope $n_s=0.966$, and a tensor-to-scalar ratio of $r = 0$.

We adopt a logarithmic binning in $k$ with $r = 2^{1/8}$, $k_{\rm min}=2\times 10^{-4}\,\mathrm{Mpc}^{-1}$, and $k_{\rm max}=0.2048\,\mathrm{Mpc}^{-1}$ (cf.\,\ref{eq: binning}) giving a total of $n_{\rm bin}=81$ one-dimensional bins. This is fine enough to ensure that both squeezed and non-squeezed configurations are densely sampled, yet coarse enough to remain practically computable. To further reduce computation time, we omit triangle configurations with $\bar{k}_3\leq 10^{-2}\,\mathrm{Mpc}^{-1}$ (for $\bar{k}_1\leq \bar{k}_2\leq \bar{k}_3$), which contribute negligible signal-to-noise (as demonstrated below). This results in $N_{\rm bin}^{\rm 3D}=5424$ triangular-configurations, which are compressed to $171$ bins in $S_{\rm 2D}(x,y)$. 

Despite the efficient \textsc{cython}-accelerated \textsc{PolySpec} code, computing the finely-binned shape function is an expensive operation. Here, computation requires $400$ node-hours for the Fisher matrix (using $20$ Monte Carlo realizations, which is more than sufficient for convergence) and $9$ node-hour per simulation for the numerators, with the execution time dominated by the linear term in \eqref{eq: 3d-estimator}. This piece can be omitted at negligible loss of optimality if one restricts to $\bar{x}\gtrsim 0.1$ (which reduces the cost to $6$ node-minutes per simulation). We stress, however, that this computation only has to be performed once per dataset (rather than once per model). \resub{Our measurements are publicly available \href{https://github.com/oliverphilcox/Binned-Bispectrum-Shape}{online}.}

\subsubsection{Comparison of Errorbars}

\begin{figure}
    \centering
    \includegraphics[width=0.8\linewidth]{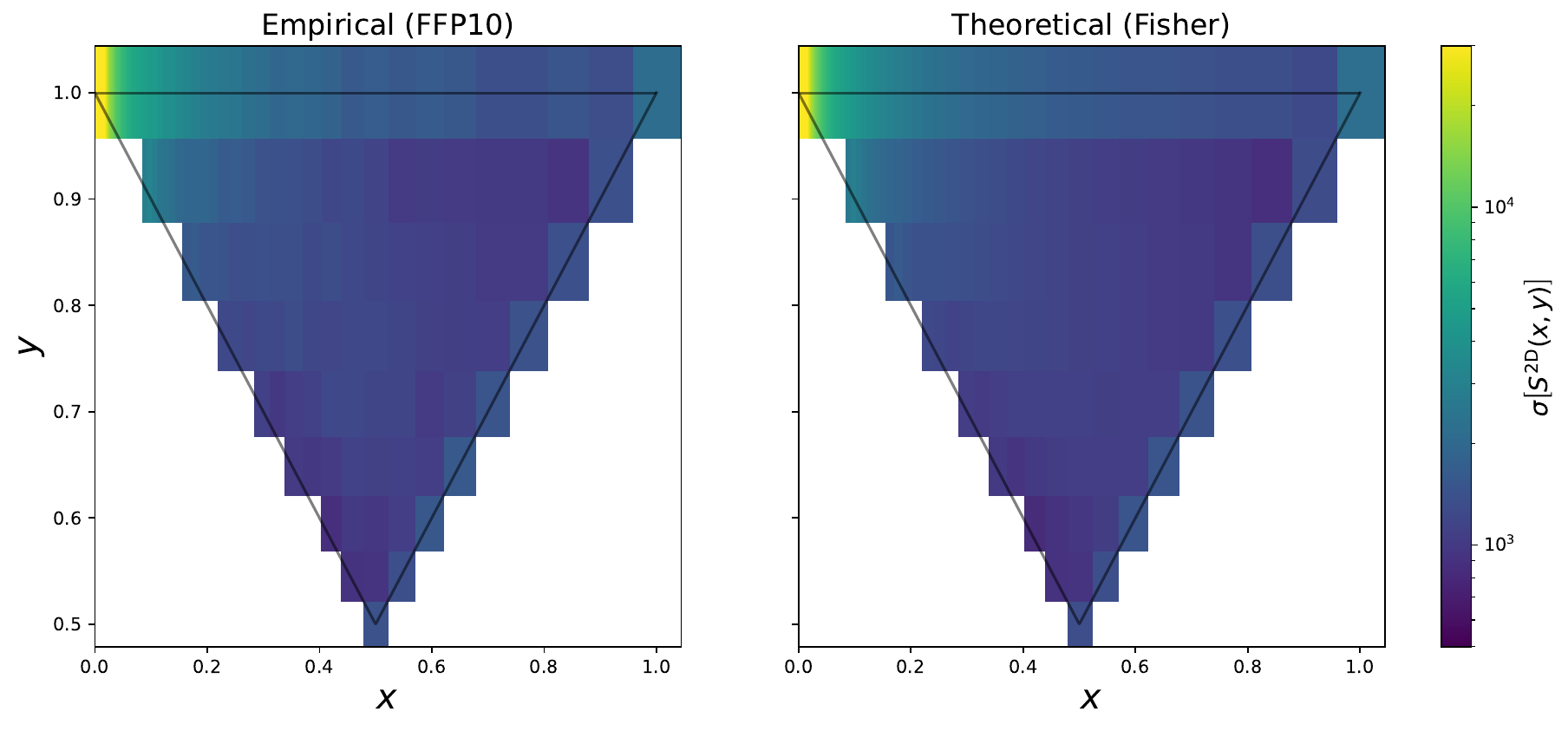}
    \caption{Comparison of the empirical (left) and theoretical (right) $1\sigma$ errors on the two-dimensional bispectrum shape. The former are obtained using $400$ FFP10/\textsc{npipe} simulations, whilst the latter are computed from the inverse of the (mask- and beam-dependent) Fisher normalization matrix. We find excellent agreement across the full two-dimensional plane, indicating that our measurements are close-to-optimal. A comparison of the errors in the squeezed-limit is shown in Fig.\,\ref{fig: error-comparison-squeezed}.}
    \label{fig: error-comparison}
\end{figure}

\begin{figure}
    \centering
    \includegraphics[width=0.5\linewidth]{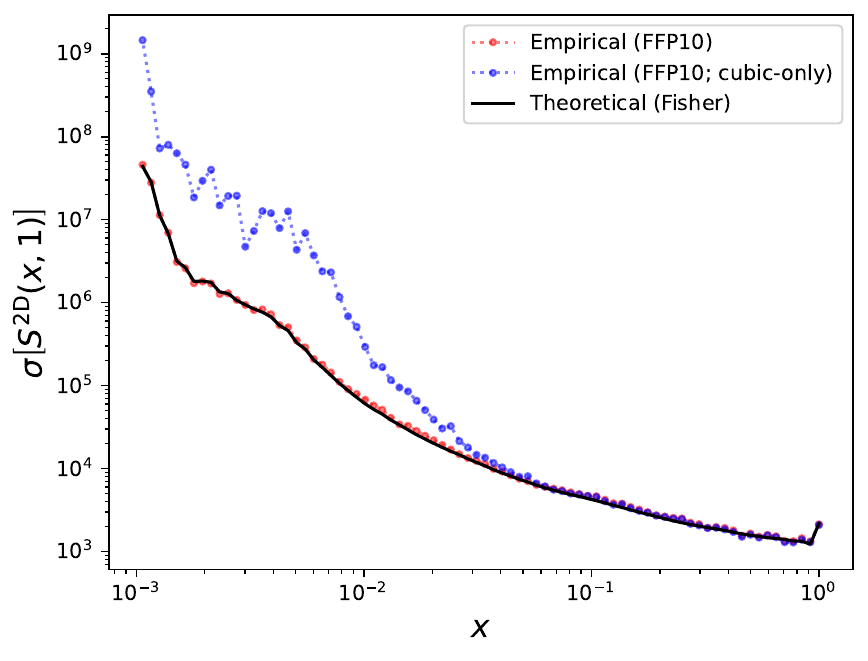}
    \caption{As Fig.\,\ref{fig: error-comparison} but restricting to isosceles triangles with $y=1$. We find excellent agreement between empirical (red) and theoretical (black) errorbars across three orders-of-magnitude in $x$, validating our implementation. We also show the errorbars obtained when dropping the linear term in the estimator (blue, \ref{eq: 3d-estimator}); as expected, this term significantly significantly reduces the variance for highly squeezed configurations. We observe that configurations with squeezing ratios beyond $x\approx 0.002$ cannot be meaningfully constrained with \textit{Planck} data.}
    \label{fig: error-comparison-squeezed}
\end{figure}

\noindent In Figs.\,\ref{fig: error-comparison}\,\&\,\ref{fig: error-comparison-squeezed}, we compare the errorbars on the two-dimensional shape coefficients obtained from the 400 FFP10 simulations and the inverse normalization matrix, $\F^{-1}$. We find excellent agreement across the full $(x,y)$-plane (with a ratio of $1.04\pm0.04$), including in highly squeezed configurations. This implies that our estimator is close-to-optimal (\textit{i.e.}\ it approaches the minimum-variance estimator for $s^{\rm 2D}$ via $s^{\rm 3D}$). As noted in the main text, the errorbars on $S^{\rm 2D}(x,y)$ scale as $x^{-1}$ as we approach the squeezed limit; moreover, the data are essentially unable to constrain extremely squeezed configurations with $x\lesssim 0.002$ (as discussed below). We also note that the errorbar is somewhat larger for isosceles and equilateral triangles: this is due to a symmetry factor (noting that there are six types of triangle corresponding to general $x,y$ but only one with $x=y=1$).
Finally, we can assess the impact of the linear term in the bispectrum estimator. For non-squeezed configurations (with $x\gtrsim 0.05$), dropping the linear term has no effect on the empirical errors (as expected), whilst for highly squeezed configurations, its excision inflates the errorbar by up to $30\times$. This indicates that the linear term is a crucial part of the squeezed shape-function estimator.

\subsubsection{Constraints on Standard Templates}
\noindent Using our shape function measurements, we can constrain a wealth of scale-invariant primordial bispectra. Practically, this can be achieved using the Gaussian likelihood for a set of model amplitudes $\{f_{\rm NL}^{(i)}\}$:
\beq\label{eq: fnl-likelihood}
    -2\log \mathcal{L}(\{f_{\rm NL}\}) &=& \sum_{NN'}\left(\widehat{s}^{\rm 2D}_N-\sum_if_{\rm NL}^{(i)}S^{{\rm theory},(i)}(\bar{x}_N,\bar{y}_N)\right)\mathrm{Cov}_{NN'}^{-1}\left(\widehat{s}^{\rm 2D}_{N'}-\sum_jf_{\rm NL}^{(j)}S^{{\rm theory},(j)}(\bar{x}_{N'},\bar{y}_{N'})\right)
\eeq
where $\mathrm{Cov}_{NN'}$ is the covariance of the binned shape function measurements, and $\{S^{{\rm theory},(i)}\}$ are the theoretical models, evaluated at the bin-centers and normalized to unity at $x=y=1$. Here, we consider several options for the covariance: (1) the inverse Fisher matrix, $\F^{-1}$; (2) the sample covariance from the FFP10 simulations; (3) a hybrid approach, combining the correlation structure of $\F^{-1}$ with the amplitude of the empirical covariance (obtained by rotating the FFP10 measurements into a diagonal basis using the Cholesky factorization of $\F$). Whilst option (2) includes any sources of suboptimality (sourced, for example, by our assumption of a diagonal weighting scheme), it is limited by the number of simulations (here $400$ for a $171$-dimensional statistic), thus we primarily adopt option (3) in this study (following a number of previous works \citep[e.g.,][]{PhilcoxCMB,Philcox:2024wqx}).

\begin{figure}
    \centering
    \includegraphics[width=0.5\linewidth]{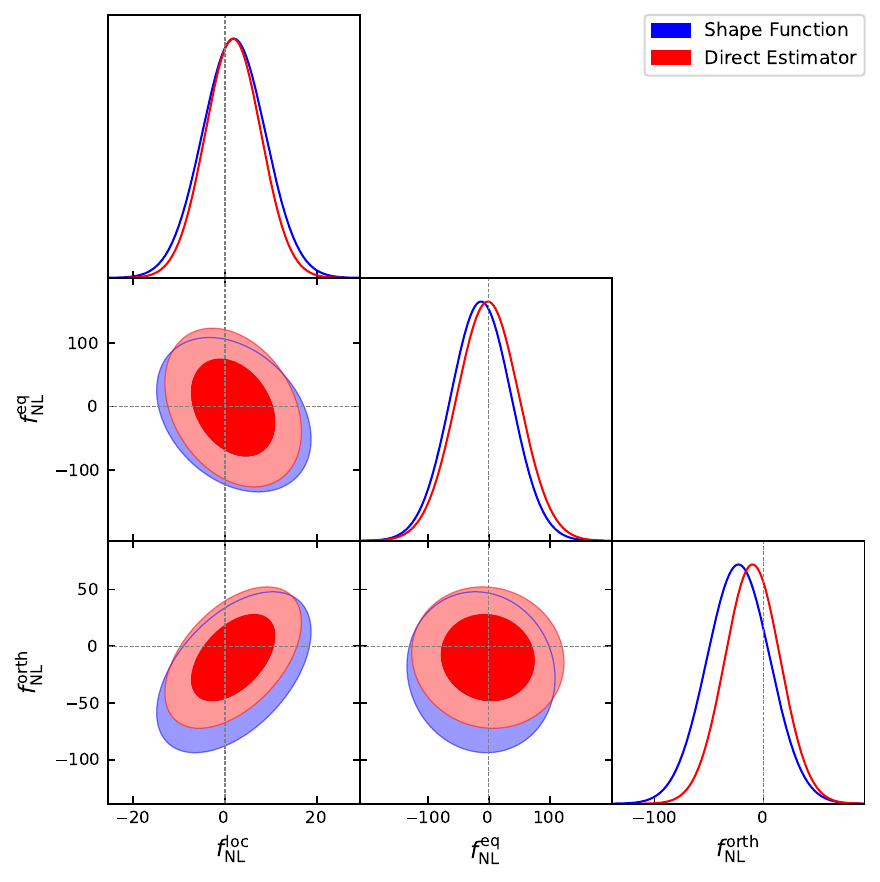}
    \caption{Joint constraints on the local, equilateral, and orthogonal bispectrum templates obtained from the shape function measurements of this \textit{Letter} (blue) and direct KSW-type estimators (red) applied to \textit{Planck} PR4 temperature and polarization data. Despite taking nine orders of magnitude less time to compute (given the binned shape measurements), the measurements derived from the reconstructed shape function are consistent with those from the optimal KSW estimators within $(0.02-0.4)\sigma$ for the best-fit and $(3-13)\%$ for the errorbar. This indicates that our compression does not lead to significant loss of information.}
    \label{fig: template-constraints}
\end{figure}

A powerful way to validate our approach is to compare the $f_{\rm NL}$ constraints obtained from \eqref{eq: fnl-likelihood} to those derived using alternative approaches. Here, we perform this test for three canonical models: local, equilateral, and orthogonal, comparing against the results from the direct KSW-style estimators included in \textsc{PolySpec} (which take the form of \eqref{eq: 3d-estimator} with $s^{\rm 3D}_n\to f_{\rm NL}$, and exploit the factorizability of the templates). 
In Fig.\,\ref{fig: template-constraints} we display the joint constraints on $\{f_{\rm NL}^{\rm loc},f_{\rm NL}^{\rm eq},f_{\rm NL}^{\rm orth}\}$ from the two methods. Overall, we find fairly good agreement, with the marginalized best-fits and errorbars consistent within $0.4\sigma$ and $13\%$ respectively. The slight loss of information in the shape-derived measurements relative to the (optimal) direct estimates is attributed to the data compression inherent in the method and could be reduced by using finer $(x,y)$-bins. Although an $\mathcal{O}(10\%)$ loss of information is not negligible, we stress that this is accompanied by a tremendous improvement in computation time; given the precomputed binned bispectrum coefficients, it takes only milliseconds to obtain $f_{\rm NL}$ constraints for an arbitrary scale-invariant model. 

We can also use these results to test the covariance and binning assumptions. Switching from the hybrid covariance to the inverse Fisher covariance (schemes (3) and (1) above) changes the best-fits and errorbars by at most $0.1\sigma$ and $5\%$, indicating good consistency. Using the sample covariance from 400 FFP10 simulations we find larger shifts in the best-fit (up to $0.9\sigma$), which we attribute to noise in the covariance (expected given the limited sample size). Replacing the theoretical models in \eqref{eq: fnl-likelihood} with those explicitly integrated across each bin (following \eqref{eq: S-2d-average}) leads to negligible changes in the constraints ($<0.1\sigma$), indicating that, for the smooth templates considered in this study, bin integration is not required.

\subsubsection{Scale Dependence}
\noindent In the main text, we have focused on the measurement of $S^{\rm 2D}$, invoking scale-invariance to combine a large number of three-dimensional measurements into two-dimensional bins. In this section, we take a different approach, using three-dimensional estimators to assess the inflationary information content as a function of scale. 
For this purpose, we use a broad $k$-space binning specified by $r=2^{1/2},\,k_{\rm min}=2\times 10^{-5}\,\mathrm{Mpc}^{-1},\,k_{\rm max}=0.32768\,\mathrm{Mpc}^{-1}$ (cf.\,\ref{eq: binning}). Rather than considering all possible triangular configurations formed from the available one-dimensional $k$-bins, we restrict to four types of shape: equilateral ($\bar{x}=\bar{y}=1$), flattened ($\bar{x}=\bar{y}=0.5$), squeezed ($\bar{x}=1/8$, $\bar{y}=1$) and highly-squeezed ($\bar{x}=1/64$, $\bar{y}=1$), defining $\bar{x}=\bar{k}_1/\bar{k}_3$, $\bar{y}=\bar{k}_2/\bar{k}_3$ as usual. Using the same set-up as before (except reducing to $100$ covariance simulations for speed), we compute the bin amplitudes using the \textsc{PolySpec} code (noting that the above restrictions are equivalent to a particular choice of projection matrix $\P$ in \eqref{eq: 2d-estimator}).

\begin{figure}
    \centering
    \includegraphics[width=0.44\linewidth]{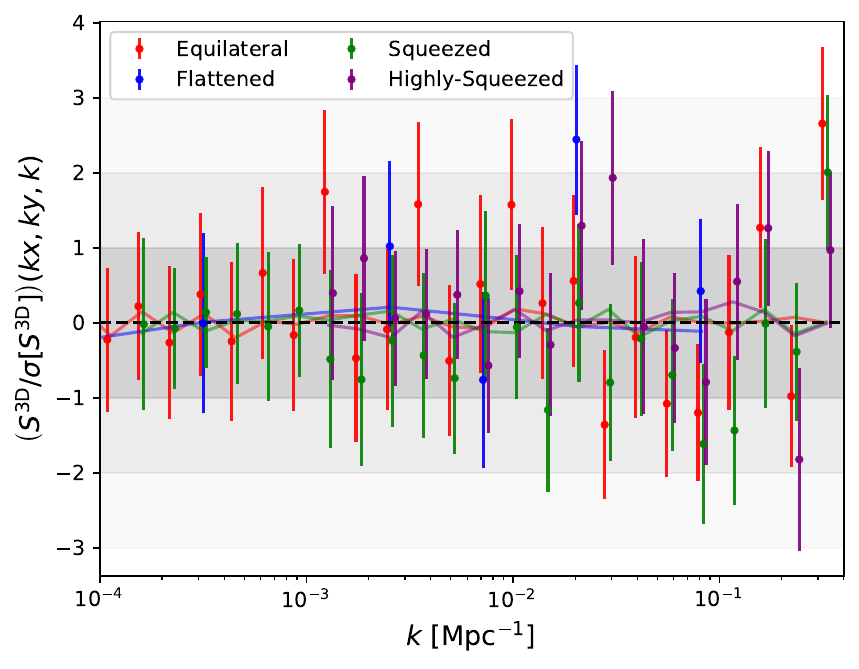}
    \includegraphics[width=0.55\linewidth]{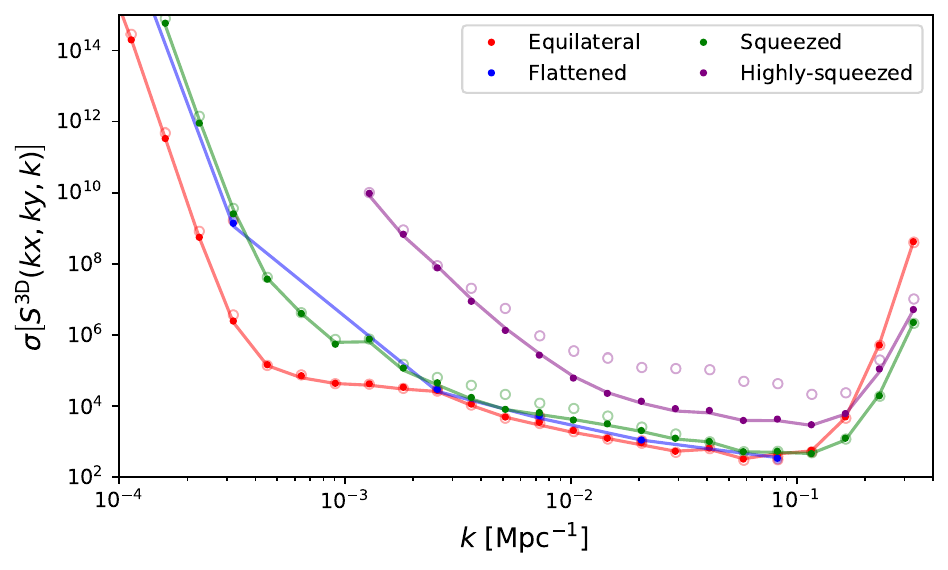}
    \caption{Constraints on the dimensionless primordial shape as a function of scale for four types of triangle configuration. \textbf{Left panel}: Signal-to-noise of the \textit{Planck} PR4 constraints. The solid lines and errorbars indicate the mean and $1\sigma$ errorbars from FFP10 simulations. We omit configurations with $kx\leq 10^{-4}\,\mathrm{Mpc}^{-1}$ which do not contain appreciable signal.
    \textbf{Right panel}: Constraining power as a function of scale. The lines (points) indicate $1\sigma$ errors obtained from the theoretical (empirical) covariances, and open circles indicate the results obtained when omittng the linear term in the estimator (which impacts only highly-squeezed configurations). Most of the information content comes from triangles with $k\in [10^{-3},2\times 10^{-1}
    ]\,\mathrm{Mpc}^{-1}$, with peak sensitivity around $k=0.1\,\mathrm{Mpc}^{-1}$.}
    \label{fig: scale-dependence}
\end{figure}

In Fig.\,\ref{fig: scale-dependence}, we show the constraints on $S^{\rm 3D}(kx,ky,k)$ as a function of the largest momentum, $k$. As in the two-dimensional analysis, we find no evidence for a non-zero signal, with a maximum deviation of $2.6\sigma$ across 96 configurations. Turning to the errorbars, we find that the CMB is insensitive to both very large and very small scales: configurations with $kx\lesssim 5\times 10^{-4}\,\mathrm{Mpc}^{-1}$ and $k\gtrsim 2\times 10^{-1}\,\mathrm{Mpc}^{-1}$ exhibit huge errorbars as well as strong correlations between neighboring bins (reaching $96\%$). In the intermediate regime, $\sigma[S^{\rm 3D}]$ scales approximately $k^{-1}$ (matching scaling arguments, noting that the CMB is a two-dimensional tracer \citep[e.g.,][]{Kalaja:2020mkq}). Regardless of the triangle configuration, we find peak sensitivity when $k\approx 0.1\,\mathrm{Mpc}^{-1}$; this corresponds to $\ell$ of around $1000$, indicating a trade-off between cosmic-variance and experimental noise. The main conclusion from this exercise is the following: the \textit{Planck} CMB is sensitive to primordial non-Gaussianity across approximately six $e$-folds of inflationary evolution, with a constraints dominated by triangles where the largest $k$ is around $0.1\,\mathrm{Mpc}^{-1}$.

\section*{Multi-Field Lagrangian}
\noindent Here, we specify the inflationary Lagrangian used to compute the exchange bispectra shown in Fig.\,\ref{fig: bootstrap-constraints}. In terms of the Goldstone mode of broken time translations, $\pi$, the basic Lagrangian is given by
\beq
    a^{-3}\mathcal{L} &\supset& \frac{M_{\rm Pl}^2|\dot H|}{c_s^2}\left(\dot\pi^2-c_s^2\frac{(\partial_i\pi)^2}{a^2}\right)+a^{-3}\mathcal{L}_\pi^{(\rm int)},
\eeq
where $H$ is the Hubble parameter, $c_s$ is the sound-speed, and $\mathcal{L}_\pi^{(\rm int)}$ represents higher-order self-interactions that we ignore in this study. Introducing a scalar field, $\sigma$ (optionally with spin, but setting $c_\sigma=1$ without loss of generality), we induce new terms in the Lagrangian:
\beq
    a^{-3}\mathcal{L} &\supset & a^{-3}\mathcal{L}_\sigma^{(2)} + 
    \begin{cases}\rho_0\dot\pi\sigma+\lambda_0\dot\pi^2\sigma\color{black}+a^{-2}\lambda_0'(\partial_i\pi)^2\sigma& \text{spin-0}\\
    a^{-2}\left(\rho_1\partial_i\pi\sigma^i+\lambda_1\dot\pi\partial_i\pi\sigma^i\right)& \text{spin-1}\\
    a^{-4}\left(\rho_2\partial_{ij}\pi\hat\sigma^{ij}+\lambda_2\dot\pi\partial_{ij}\pi\hat\sigma^{ij}\right) & \text{spin-2},\end{cases}
\eeq
where $\mathcal{L}_\sigma^{(2)}$ is the free-field quadratic Lagrangian. We have dropped terms that contribute only to double- and triple-exchange diagrams (e.g., $\dot\pi\sigma^2$), though these can source large interaction signatures \citep[e.g.,][]{Kumar:2026ogn}. In the main text, we constrain the $f_{\rm NL}$ amplitudes corresponding to each of the four cubic interactions above, with $f_{\rm NL}\propto \rho_n\lambda_n$. \resub{Note that we do not impose perturbativity constraints on $f_{\rm NL}$: these will be discussed in \citep{cosmoflow2}.}

\bibliographystyle{apsrev4-2}
\bibliography{refs}

\end{document}